# The mass of the Higgs-like boson in the four-lepton decay channel at the LHC

V. Roinishvili

*JINR, Dubna, Russia and EAIP, Tbilisi, Georgia*

ABSTRACT

*A very simple and transparent way for the mass definition of a new boson, probably Higgs (H), observed at LHC, decaying into 4 leptons, is presented. The obtained mass of H is 125.5±0.4 GeV with today statistics.*

INTRODUCTION

In the resent publication on H→4 leptons the ATLAS [1] and CMS [2] collaborations give the following values for the measured mass of H in this decay channel:

$$124.3\pm0.5(st)\pm0.6(syst) \text{ GeV – ATLAS and } 125.8\pm0.5(st)\pm0.2(syst) \text{ GeV – CMS} \quad (1)$$

Fig. 1 shows the distribution of the effective mass of 4 leptons-$M_{4l}$ for ATLAS and CMS extracted from the corresponding figures in [1] and [2]. The 4-lepton samples have the advantage that they contain two signals: one from H under the study and another from the very well known Z boson. The last can be used to obtain the experimental resolution and a possible systematic shift of the 4-lepton effective mass scale in the *selected* samples without MC simulations.

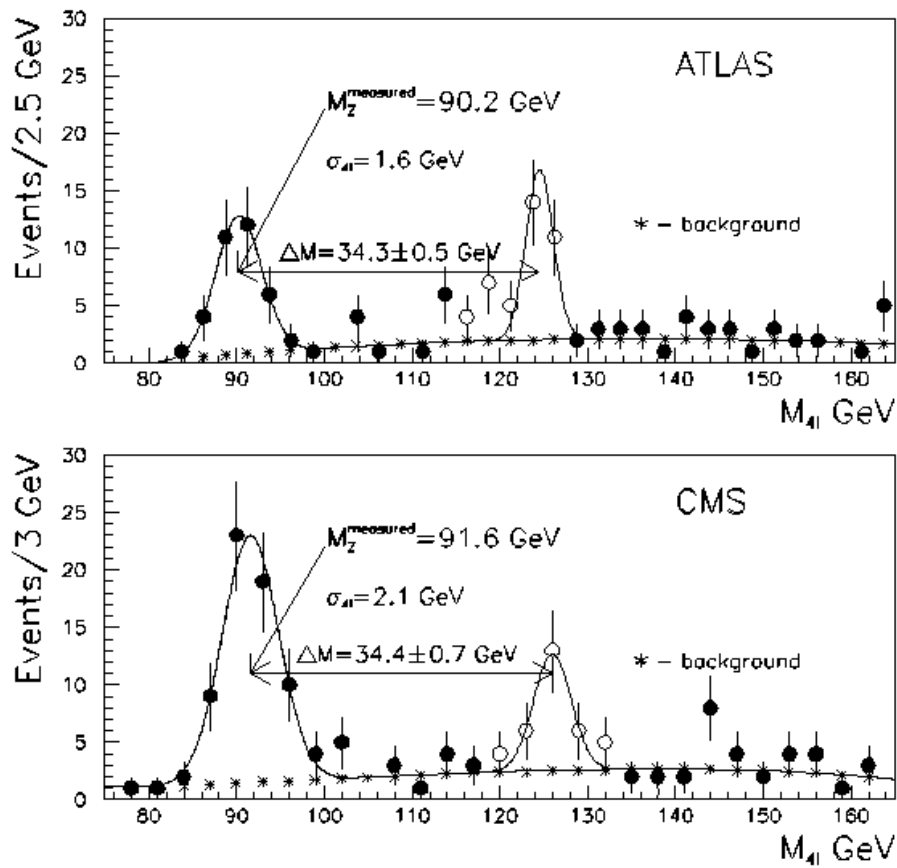

Fig. 1   The distributions of the four-lepton effective mass

The main idea of the present approach is to determine experimentally the mass difference between H and Z – ΔM, and then to define the mass of H as the sum of ΔM and the table value of the Z mass – $M_Z$(table)=91.19 GeV. In this case the constant systematic shifts of the mass scale will be cancelled. Such a definition of the mass leads to a surprisingly good coincidence between ATLAS and CMS as it will be shown below.

ANALYSIS AND RESULTS

The distributions on Fig. 1 were fitted (MINUIT) in the range of the effective mass 4 leptons - $M_{4l}$ 75 < $M_{4l}$ < 165GeV without white points to define experimentally backgrounds – *BKG* and 4-leptons effective mass resolutions – $\sigma_{4l}$ by the sum of Gaussian – $G_Z$, describing the signal of Z and the 3$^{nd}$ order polynomial describing the background:

$$Nevents = G_Z + BKG = \frac{N_Z \times bin}{\sqrt{2\pi}\sigma_Z}\exp[-\frac{(M_{4l}-M_Z)^2}{2\sigma_Z^2}] + BKG, \quad \sigma_Z = \frac{\Gamma_Z}{2\times 1.177} + \sigma_{4l} \quad (2)$$

Here $N_Z$ is the integrated number of Z events, *bin* is the binning of the abscissa (2.5 GeV for ATLAS and 3 GeV for CMS), $\Gamma_Z$=2.5 GeV is the table value of the full widths of Z and $\sigma_{4l}$ is the experimental resolution of $M_{4l}$. Free parameters were $N_Z$, $M_Z$, $\sigma_{4l}$ and the coefficients of the polynomials. For the $M_Z$ and $\sigma_{4l}$ the following values were obtained:

$$M_Z^{mesuared} = 90.2 \text{ GeV} - \text{ATLAS and } M_Z^{mesuared} = 91.6 \text{ GeV} - \text{CMS,} \quad (3)$$

$$\sigma_{4l} = 1.6 \text{ GeV} - \text{ATLAS and } \sigma_{4l} = 2.1 \text{ GeV} - \text{CMS} \quad (4)$$

After that the distributions on Fig. 1 were fitted including the points of H (white points) by the sum $G_Z + G_H + BKG$, where:

$$G_H = \frac{N_H \times bin}{\sqrt{2\pi}\sigma_H}\exp[-\frac{(M_{4l}-M_Z-\Delta M)^2}{2\sigma_H^2}], \sigma_H \equiv \sigma_{4l} \quad (5)$$

with fixed $G_Z$ and *BKG* and free $N_H$ and ΔM. The obtained values of ΔM are:

$$\Delta M = 34.3 \pm 0.5 \text{ GeV} - \text{ATLAS and } \Delta M = 34.4 \pm 0.7 \text{ GeV} - \text{CMS} \quad (6)$$

giving the following values for the mass of H defined as $M_H = M_Z$(table) + ΔM:

$$M_H = 125.5 \pm 0.5 \text{ GeV} - \text{ATLAS and } M_H = 125.6 \pm 0.7 \text{ GeV} - \text{CMS} \quad (7),$$

which are much more close to each other than the published results (1).

The weighted mean from the two experiments is now:

$$M_{H\rightarrow 4l} = 125.5 \pm 0.4 \text{ GeV} \quad (8)$$

DISCUSSION

It is obvious that one day the data from ATLAS and CMS will be combined for analysis. But even now it can be done by:

a) choosing the appropriate abscissa for both experiments in the form of

$$M_{4l}^{corrected} = M_{4l}^{measured} + M_Z^{table} - M_Z^{measured} \quad (9)$$

in order to take into account the systematic constant shift of measured mass of 4 leptons in the *used* samples, and

b) choosing as an ordinate the ratio of signal to the background:

$$R = \frac{Signal}{Bakground} = \frac{Nevents}{BKG} - 1 \qquad (10)$$

in order to take into account the difference of the ATLAS and CMS acceptances. Such a plot is presented on Fig. 2.

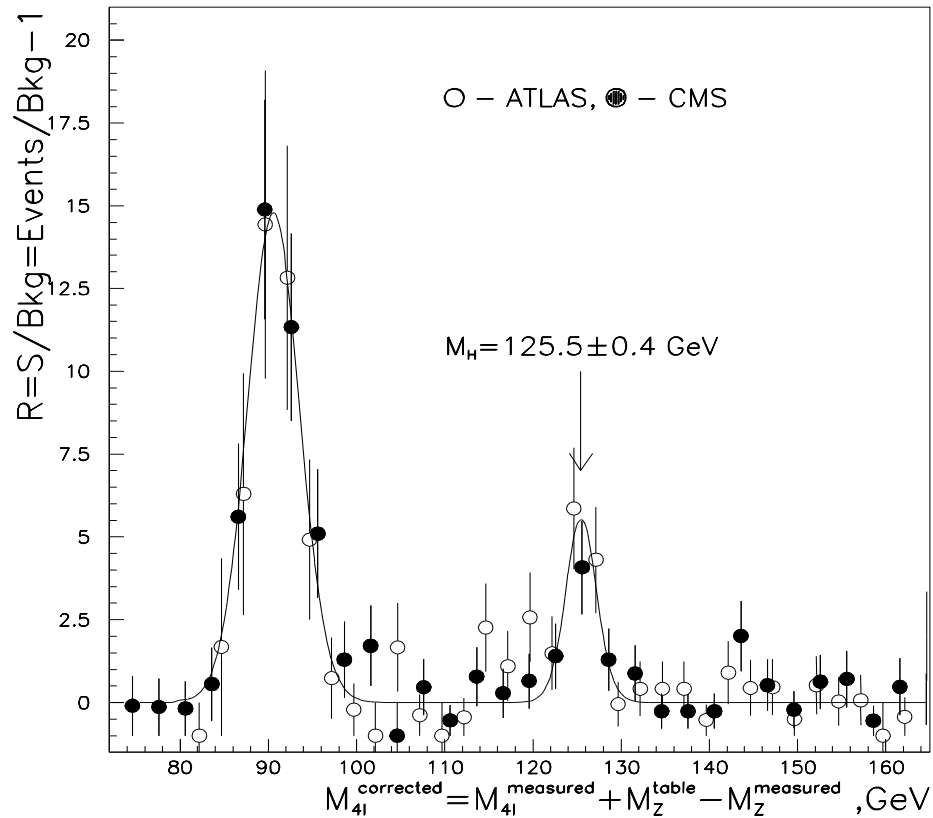

Fig. 2    The combined ATLAS and CMS ratio of signals to the background as a function of corrected four-lepton effective mass

The curve is the result of the fit of R with 3 free parameters – weights of the Z and H signals and the mass of H. The background and resolution were fixed to the ones of ATLAS and CMS correspondently. The coincidence between the two experiments is so good that one can consider the result as a result of a *common* LHC experiment with the total luminosity about twice more than accepted by ATLAS and CMS separately and which gives for the mass of H→4*l* the value of 125.5±0.4 GeV.

CONCLUSION

The measured difference ΔM between the masses of H and Z is almost free from the systematic constant shift of the mass scale which, of course, could be different for ATLAS and CMS. This allows to get a rather good coincidence between the two experiments if one will use the definition of the mass of H as the sum of the table value of the Z mass and ΔM. Obtained in this way the mass of the new boson discovered by ATLAS and CMS in the 4-lepton decay channel equals 125.5±0.4 GeV.

ACKNOWLEDGEMENTS

Author wishes to thank I. Mandjavidze from IRFU, CEA Saclay for useful discussions and help.